\documentclass[onecolumn,showkeys,amsmath,amssymb]{revtex4}
\usepackage{graphicx}
\usepackage{graphics}
\usepackage{subfigure}
\usepackage{dcolumn}
\usepackage{bm}
\usepackage{txfonts}
\usepackage{enumerate}
\usepackage{amsmath,empheq,mathrsfs}
\usepackage{multirow, booktabs}

\makeatletter
\newcommand*{\rom}[1]{\expandafter\@slowromancap\romannumeral #1@}
\makeatother
\begin{document}


\title{Asymmetric Partisan Voter Turnout Games}

\author{Cameron Guage$^{1,2}$}
\email{Cameron.J.Guage.22@dartmouth.edu}
\author{Feng Fu$^{1,2,3}$}
\email{feng.fu@dartmouth.edu}
\affiliation{$^1$ Department of Mathematics, Dartmouth College, Hanover, NH 03755, USA\\
$^2$ Program in Quantitative Social Science, Dartmouth College, Hanover, NH 03755, USA\\
$^3$ Department of Biomedical Data Science, Geisel School of Medicine at Dartmouth, Lebanon, NH 03756, USA}

\date{\today}

\begin{abstract}
    Since Downs proposed that the act of voting is irrational in 1957, myriad models have been proposed to explain voting and account for observed turnout patterns. We propose a model in which partisans consider both the instrumental and expressive benefits of their vote when deciding whether or not to abstain in an election, introducing an asymmetry that most other models do not consider. Allowing learning processes within our electorate, we analyze what turnout states are rationalizable under various conditions. Our model predicts comparative statics that are consistent with voter behavior. Furthermore, relaxing some of our preliminary assumptions eliminates some of the discrepancies between our model and empirical voter behavior.
\end{abstract}

 \maketitle

\section{Introduction}

\par A problem that has plagued political scientists for decades without satisfactory resolution is the apparent irrationality of voting. One should only vote if the benefits outweigh the costs, and, if voting is considered to be an instrumental means to an end of effecting political change, this is rarely the case~\cite{downs1957economic}. In the discussion of this anomaly, the payoff an individual receives for voting is typically represented by the following expression:
\vspace{3mm} 
\par $pB-c$, where
\par $p$ = the probability that one'€™s vote will be pivotal,
\par $B$ = the benefit differential one derives from one'€™s preferred candidate gaining office, and
\par $c$  = all opportunity costs associated with the act of voting
\vspace{3mm} 

\par Individuals only choose to vote if this expression is positive, an outcome that is extremely unlikely when considering the remarkably small values of $p$ inherent in most elections. For an American presidential election, this value is less than one in ten million~\cite{tullock1967toward}. Furthermore, in an empirical study on U.S. Congressional and state elections, it was found that only one out of every 100,000 votes cast in U.S. elections was cast for a candidate that either tied or won by a single vote~\cite{mulligan2003empirical}. The fact that $p$ is likely to be so miniscule, then, should deter any individual with even the most negligible voting costs from going to the polls, driving turnout rates effectively to zero. Paradoxically, a wise individual would realize that, with no other rational individuals voting, $p$ would effectively become one, and she (along with all other individuals) should vote, again decreasing $p$, and so on.

\par The seeming irrationality of one of the most fundamental acts of democracy set many political scientists and economists into motion in an attempt to explain how rational voters can vote with high turnout rates, even when facing non-negligible costs. Initially,~\cite{riker1968theory} offered the potential solution that Downs's model was incomplete, and that there should be another benefit term that is not scaled by the probability that one's vote is pivotal. This term represents a consumption benefit one derives from voting, and could be seen as the satisfaction one receives from fulfilling one's \emph{civic duty} via electoral participation. Riker and Ordeshook explain that adding a variable (they use $D$ in their payoff equation) for the consumption benefit one derives from voting accounts for the rationality of high voter turnout when it is clear that the probability that an individual's vote will impact the outcome of the election is staggeringly small. Many criticized this model for offering little insight into voting motivations, as the magnitude of one's €œ$D$€ term seems to almost entirely dictate the decision of whether or not to vote. Additionally, it does not account for behavior like changes in turnout in the same region for different types of elections~\cite{green1996pathologies}. Lastly, while the results from this model compare well with election data within a given year, they do not seem to hold up when analyzing election data across different years~\cite{aldrich1976some}.

\par Another proposed solution to Downs' paradox is that, rather than utility maximizers, rational individuals are regret minimaxers; that is, they choose the strategy that minimizes the chance of ending up with the result that would produce their maximum regret~\cite{ferejohn1974paradox}. While supporting high turnout equilibria, the minimax regret model has been largely discounted in the literature, with critics pointing out that a true regret minimaxer would be so risk-averse that they would never cross the street, even if the polling location were on the other side~\cite{dhillon2002economic}.

\par Others have used a game-theoretic to help explain the phenomenon of voting.~\cite{ledyard1984pure} posits a model of voting behavior that includes voters $and$ candidates as players in a voting game, but finds zero turnout in equilibrium. One of the more promising attempts to break the paradox of not voting was undertaken in~\cite{palfrey1983strategic}. Palfrey and Rosenthal find that, in their model, high turnout behavior can be supported, even in the presence of high costs. Two years later, they build on this model by introducing uncertainty about voting costs, noting that the equilibria found in their 1983 paper were fragile insofar as they rested on the assumption that costs were common knowledge to the entire electorate. They find that, once some information about voting costs is restricted to individuals themselves, the high turnout equilibria vanish~\cite{palfrey1985voter}. 

\par Others have attempted to use information in a different context to explain voting, referring to the information one has about the candidates and the potential political consequences of their elections to office. In an attempt to explain voting patterns,~\cite{matsusaka1995explaining} claims that the more information a citizen has, the higher the payoff she receives from voting is, as she is more confident in her vote. Similarly believing in the power of information to explain voter turnout, Feddersen and Pesendorfer create a model that assumes asymmetric information in a population; however, it is in discord with both Matsusaka'€™s formulation of the information effect on turnout and simple logic, as abstention rates are at times positively correlated with the proportion of informed voters in the population~\cite{feddersen1999abstention}.

\par A different voting model that has gained traction in the political science community as a potential solution to the paradox of not voting assumes that, rather than being entirely self-interested when deciding whether or not to vote, individuals are rule utilitarians~\cite{harsanyi1982rule}. Rule utilitarians follow a rule (for voting, in this setting) that, if followed by everybody, leads to the result that yields the maximum utility. Harsanyi does not indicate how this rule applies to situations in which there are divergent opinions regarding what the most socially desirable outcome is, which is undoubtedly the case when it comes to partisan politics.~\cite{feddersen2006theory} considers this possibility and allows disagreement about what the utility-maximizing outcome is, outlining a model where turnout is motivated by disagreement within an electorate. This type of model is corroborated by empirical evidence, with the group rule-utilitarian model explaining nearly half of the variation in voter turnout in Texas liquor referenda~\cite{coate2004group}. Contrary to this support,~\cite{merlo2013external} finds that, upon comparing this `ethical'€ voting model with many others using the concealed parameter recovery method, the ethical model performs relatively poorly. Furthermore,~\cite{feddersen2004rational} claims that group-based models such as the rule-utilitarian and altruistic voting model~\cite{margolis1984selfishness} are problematic as they do not ensure the existence of equilibria, allow for mixed strategy equilibria, or explain why people join groups in the first place, which Feddersen believes is relevant in the calculus of voting for group-based models.

\par Others have attempted to model voters as learners in a survival approach rather than rational utility maximizers.~\cite{sieg1995evolutionary} assumes voters repeat strategies that induced pleasure and avoid strategies that induced punishment. Sieg and Schulz find that, while in some scenarios their model predicts Downs' troubling result of zero turnout, it can also lead to individuals œ`learning€' participation.~\cite{palfrey1984participation} supports the promising nature of this work, stating that learning over time is a helpful tool for narrowing down the multiplicity of equilibria that may arise in voting games. Furthermore, evolutionary processes in electoral settings are supported by empirical studies such as~\cite{rosenthal1973electoral}, which finds learning processes are present within the electorate of the French Fifth Republic. While Sieg and Schulz'€™s model certainly sheds light on turnout behavior, it does not allow for any mixed equilibria, a possibility that is worth considering when we observe similar people behaving in divergent ways in electoral settings.

\par Many others have conducted experiments and empirical studies to examine voter behavior and test the validity of some of the different voter models proposed in the literature. Some have studied the effect of different representation systems on turnout~\cite{schram1996voter};~\cite{blais1990does}, while others have focused on the comparative statics of election data to observe how factors such as size and perceived closeness relate to turnout. Many studies have found a \emph{size effect,} wherein turnout decreases as electorate size increases~\cite{romer1983voting};~\cite{hansen1987downsian};~\cite{barzel1973act};~\cite{breux2017fewer};~\cite{levine2007paradox} and many have found a \emph{closeness effect,} wherein turnout decreases as the closeness of an election decreases~\cite{thompson1982closeness};~\cite{levine2007paradox}, although closeness can be evaluated in different ways.~\cite{kaniovski2006community} tries to explain these relationships qualitatively, claiming that electoral closeness and size reflect heterogeneity in an electorate, which in turn increases voter turnout. Other studies use these results as evidence that voters are rational actors that follow Downs'€™ voting utility formulation, as their votes are more likely to be pivotal in small electorates and close elections. The consensus seems to be that, while this evidence supports the idea that voting is not solely motivated by a `€œconsumption benefit' that one invariably gets from going to the polls, it is not solely motivated by pivotal-vote considerations either. For example, Breux et al. finds that rational choice theory can explain approximately 45\% of voter turnout in municipal elections. An empirical study of presidential elections found that, while the rational voter hypothesis seems unable to explain turnout in its entirety, neither can `€œcivic duty' arguments ~\cite{foster1983performance}.~\cite{green1996pathologies} also notes that rational utility maximization is just a $part$ of the explanation for why people vote.

\par In fact, there is a growing literature against $homo$ $economicus$, or the idea that human behavior, and in this case voting behavior, should be viewed strictly through the economic lens of utility maximization that Downs placed on it. This is not to say that voters are irrational, but rather that they have payoff structures that reflect considerations Downs did not include. One can then analyze how rational individuals would act in this new voting setting that is more reflective of political considerations.~\cite{overbye1995making} challenges $homo$ $economicus$ by blending economic and sociological concepts when analyzing voter behavior. Overbye conceptualizes voting as an investment in the reputation that one is concerned about the public good.~\cite{brennan1997democracy} points out that political and economic behavior are inherently different, with political behavior having not only the instrumental benefit that an economic decision has, but also an expressive benefit.~\cite{brennan2008psychological} expands upon this idea by explaining that, while people certainly try to affect outcomes when they vote, they also benefit from expressing their political views. Brennan explains that humans exhibit expressive behavior all the time, from going to watch one'€™s favorite team play football to sending a get well soon card to a relative. Importantly, he notes that expressive behavior is not necessarily outcome-independent, which would have an effect similar to that of the Riker-Ordeshook model's \emph{$D$} term. With this idea in mind, we posit a game-theoretic model for voting behavior that blends instrumental and expressive motivations in an attempt to better understand the dynamics behind voter turnout in partisan elections. We outline our base model in Section 2. In Section 3, we analyze this model and its implications for voter behavior. Section 4 compares the turnout predictions of our model with trends in real electoral data. Section 5 discusses some possible extensions to our model, and in Section 6 we conclude.

\section{Base Model}
\par Our model assumes a two-candidate election in which each member of the electorate has preferences regarding which candidate's election to office they perceive will lead to a superior outcome. The electorate can then be partitioned into two blocs, one that supports the first candidate, which we will henceforth call candidate $A$, and one that supports the second candidate, which we will refer to as candidate $B$. We further assume that individuals will not vote for the candidate they do not support, a finding consistent with~\cite{herzberg1988results}, which claims that citizens tend to vote sincerely. For simplicity, we assume that more people support candidate $A$ than candidate $B$.  We consider an $N$ person election in which $p_A$ ($>00.5$) is the proportion of the electorate that supports candidate $A$ and $p_B$ (= $1 - p_A$) is the proportion of the electorate that supports candidate $B$. We assume a simultaneous-vote and winner-take-all (majority-rule) election, lest pivotal concerns become irrelevant, as they could in a system of proportional representation. Each individual in our model then faces a binary decision: vote for her candidate of choice, or abstain. Furthermore, there are three distinct electoral outcomes: one'€™s candidate of choice wins, one'€™s candidate of choice loses, or one's candidate of choice ties. With this in mind, we define the payoff structure for a supporter of candidate $j \in{(A, B)}$ as follows:
\begin{equation}
\begin{cases}
    -1 & \text{Vote and candidate $j$ loses} \\
    \ 1-c(p_j - 0.5) & \text{Vote and candidate $j$ wins} \\
    \ 1/2 & \text{Vote and candidate $j$ ties} \\
    \ 0 & \text{Abstain} \\
    \end{cases}
\end{equation}

\vspace{3mm} 
\par To provide rationale for these payoffs, we urge the reader to participate in a thought experiment. In a similar vein to~\cite{brennan2008psychological}, our quasi-expressive voting model can be compared to the expressive act of seeing one's favorite football team play. When considering whether or not to go to the game, one weighs the potential benefits with the costs of attendance, such as gas, tickets, and the opportunity cost of foregone leisure. If one's preferred team is going to lose, one is better off staying at home and not incurring the costs of attendance. That is, there is no expressive benefit to attending the game if one's team of choice loses; nobody wants to walk out of the stadium wearing the colors of the losing team, and one could have done better by staying home. 
\par However, if one's team is going to win, one will usually want to attend the game. It is not enough to just be a supporter of the winning team; if one wants to reap the benefits associated with victory, be it the high fives with other fans or storming the field upon victory, one must have gone to the game. In this case, one is glad to have expressed one's preferences, even with the knowledge that one's cheering likely did not affect the outcome of the game. Nonetheless, there is an expressive benefit for showing up to contribute to one's favorite team when this team wins. 
\par Furthermore, we believe that the utility one gets from attending a game in which one's team is victorious is affected by how much one was expecting one's team to win. One can imagine a game in which team $A$ is almost sure to beat team $B$. The joy imparted to an $A$ fan when they nearly inevitably win would certainly be different than the joy imparted to a $B$ fan who risked coming to the game and saw a shocking underdog victory over the mighty team $A$. Thus, our model includes what we call an `underdog consideration,' wherein the majority's winning payoff is decreasing in the magnitude of the majority, and the minority's winning payoff is increasing in the magnitude of the majority. In our case, $p_j$ is a proxy for individuals' perceptions about the likelihood that candidate $j$ will win the election. The $c$ variable is a non-negative coefficient (an assumption we will relax later) that scales the degree to which a particular electorate factors this underdog consideration into its payoff structure.
\par If one's favorite team ties, the game goes into overtime; that is, a sure loss is prevented. In electoral politics, there is a surprising lack of consistency regarding the resolution of ties, with tiebreakers like coin flips, re-elections, and, in the 2020 Iowa caucus, even card draws deciding the fates of the candidates. Regardless, a sure loss is almost always prevented when an electoral tie occurs. It is noteworthy that, in our payoff structure, the payoff for voting when the election results in a tie is always positive, and in fact can exceed the payoff for voting when one's candidate of choice wins. In our football analogy, one might imagine that, however unlikely, one's presence at a game that ends in a tie may have played a role in preventing one's favorite team from losing the game: that if one less voice had been cheering, the team might just have lost the game (from a lack of motivation or something of the sort). One's role in propping up this victory is what motivates this payoff. If this seems far-fetched, consider the electoral context. In the case of a tie, it is not an exaggeration to think that one's vote prevented one's candidate of choice from losing the election; in fact, one would err in saying otherwise. Not only is the individual we are considering pivotal in preventing a loss, but also so is every individual who voted for the same candidate, as well as all of the individuals who voted for the other candidate, resulting in no negative payoffs in the electorate if this is the electoral outcome. While voting is certainly an expressive action, it would be ignorant to assume that individuals do not consider the instrumental impact of their vote, and that they would not derive benefit from playing a pivotal role in preventing the electoral defeat of their preferred candidate.
\par Now, when it is time to decide whether or not to go to the football game, the fans of each team (or, as one can likely infer by now, the supporters of each party) now face a decision: given the possible outcomes and one'€™s respective preferences over these outcomes, is leaving the costless but benefit-less comfort of home worthwhile? 
\par If we weigh the potential payoffs for voting by their respective likelihoods of occurring and sum over the three possible cases, we can obtain the expected payoff for voting. If $V_j$ is the number of individuals that vote for candidate $j$, then the payoff an $A$ supporter expects to receive for voting is

\begin{equation}
P(V_A  > V_B)[1-c(p_A-0.5))] + P(V_A  < V_B)(-1) + P(V_A  = V_B)(1/2)
\end{equation}

And the payoff a $B$ supporter expects to receive for voting is
\begin{equation}
P(V_A  > V_B)(-1) + P(V_A  < V_B)[1-c(p_B-0.5)] + P(V_A  = V_B)(1/2)
\end{equation}
\par Comparing this expected payoff to the sure zero payoff for abstaining, individuals will choose to vote when the expected payoff for voting is greater than zero and abstain when it is less than zero.
\par As all citizens face the binary choice of vote vs.\ abstain, we can model the number of individuals who turn out to vote for each party as a binomial random variable. Thus, while we assume symmetry within parties, there is an asymmetric aspect to our game, insofar as voters from different parties can decide to vote with different probabilities. If supporters of candidate $j$ will vote with probability $q_j$, then
\vspace{3mm} 
\par $V_A \sim binom(Np_A, q_A)$, and
\vspace{3mm} 
\par $V_B \sim binom(Np_B, q_B)$
\vspace{3mm}
\par As the probability that candidate $A$ wins the election is $P(V_A  > V_B)$, this can be expanded to
\vspace{3mm}
\par
$$\sum_{k=0}^{Np_B}\binom{Np_B}{k}(q_B)^k(1-q_B)^{Np_B-k}\sum_{j=k+1}^{Np_A}\binom{Np_A}{j}(q_A)^j(1-q_A)^{Np_A-j}$$
\vspace{3mm}
\par Similarly, the probability that candidate $B$ wins is
\vspace{3mm}
$$\sum_{k=0}^{Np_B-1}\binom{Np_A}{k}(q_A)^k(1-q_A)^{Np_A-k}\sum_{j=k+1}^{Np_B}\binom{Np_B}{j}(q_B)^j(1-q_B)^{Np_B-j}$$
\vspace{3mm}
\par And the probability of a tie is
\vspace{3mm}
$$\sum_{k=0}^{Np_B}\binom{Np_A}{k}(q_A)^k(1-q_A)^{Np_A-k}\binom{Np_B}{k}(q_B)^k(1-q_B)^{Np_B-k}$$
\vspace{3mm}
\par In order to analyze equilibrium behavior in our model, we plot the points at which $A$ and $B$ supporters are indifferent between voting and abstaining (that is, when their expected payoffs for voting are exactly equal to zero) over $q_A$ and $q_B$ to see what turnout states are rationalizable.
\par We find that, upon varying our three main parameters ($N$, $p_A$, and $c$), these indifference functions can be configured in different ways that have different implications for voter behavior. We go on to analyze the effects of changing these variables on voter behavior in the next section.

\section{Analysis}
\par In each plot we show the curve along which $A$ supporters are indifferent between voting and abstaining (red) and the curve along which $B$ supporters are indifferent between voting and abstaining (blue). The x-axis of each graph is $q_A$ and the y-axis is $q_B$, with both of these variables taking on values in the range [0,1]. Figure 1 shows the effect of varying $N$ and $p_A$ in our model when we set $c$ equal to zero.
\begin{figure}[htp]
\centering
\includegraphics[width=12cm]{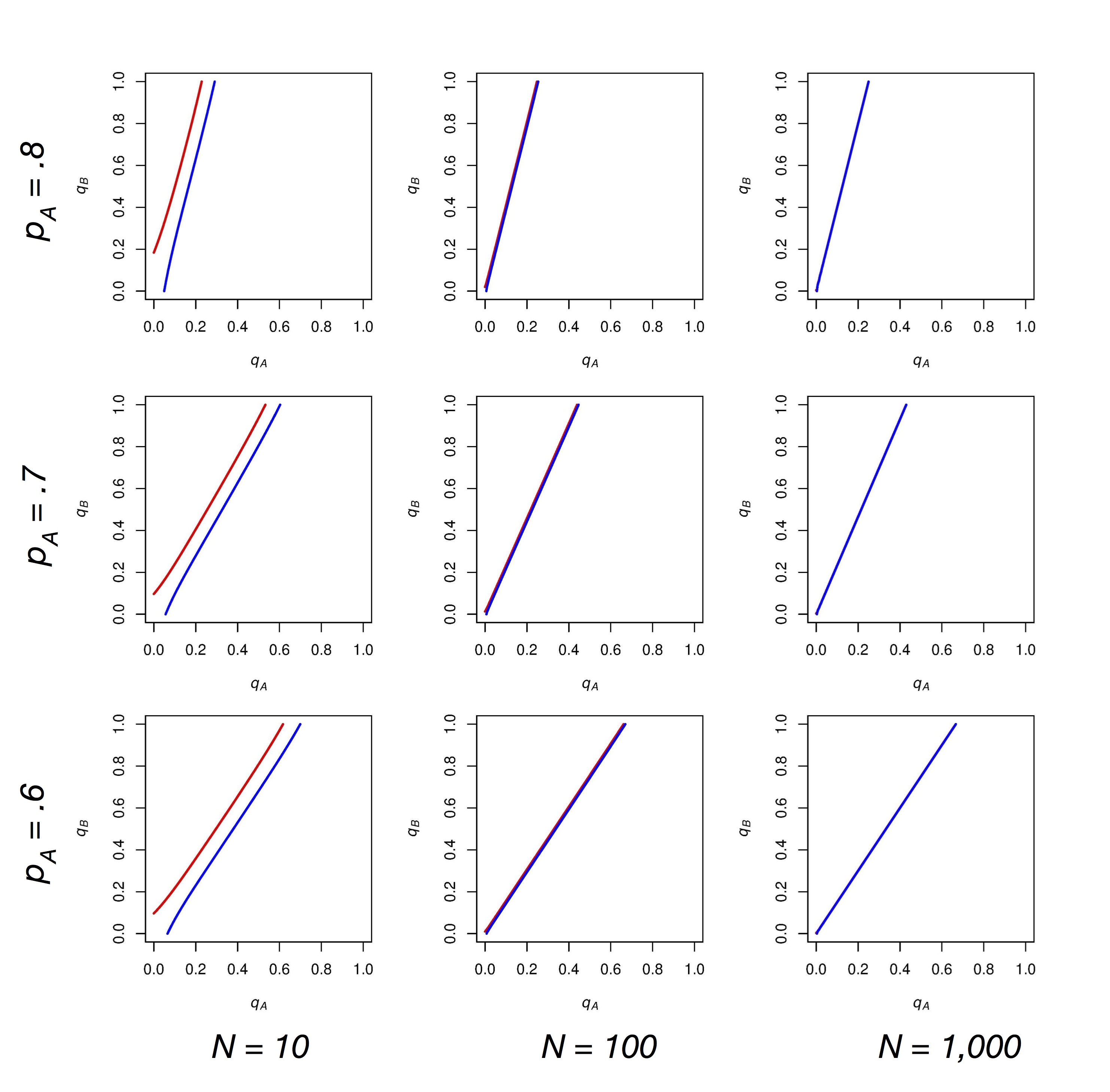}
\caption{The effect of varying $N$ and $p_A$ on the indifference functions of $A$ (red) and $B$ (blue) supporters while $c = 0$. These functions never intersect, but approach collinearity as $N \rightarrow \infty$.}
\label{Fig1}
\end{figure}

\par As we let $N$ grow, $A$ and $B$'s indifference curves converge to a single line with slope $p_A/p_B$. Let us consider why this is the case. As $N$ gets large, the probability of a tie goes to zero, all else equal. Furthermore, with $c$ (and thus the impact of the underdog effect) equal to zero, this game simplifies to a game akin to the following: Abstain and get a payoff of zero, vote and get a payoff of 1 if your candidate of choice wins and -1 if your candidate of choice loses. So, supporters of a given party are indifferent between voting and abstaining (their expected payoff for voting is equal to zero) when their likelihoods of winning and losing are approximately equal, or when the expected number of $A$ and $B$ voters are close to the same. As the number of $A$ and $B$ voters are binomial variables, the expected number of $A$ voters is $Np_Aq_A$, and the expected number of $B$ voters is $Np_Bq_B$. We find that these are equal when $q_A = p_B$ and $q_B = p_A$, making the expected number of voters for both parties equal to $Np_Ap_B$. With $q_A$ and $q_B$ set to these levels, the slope ($\Delta q_B/\Delta q_A$) approaches $p_A/p_B$. With the probability of a tie approaching zero as we let $N$ get large, the indifference curves of $A$ and $B$ supporters approach the same line with this slope.
\par Now we want to analyze what turnout states are rationalizable for a given set of parameters. In order to do so, we search not just for Nash equilibria, but for $evolutionarily$ $stable$ $strategies$ (which we will sometimes refer to as $ESS$'s). Similar to Sieg and Schulz (1995), we consider the learning individual in a voting context, wherein actions that induced pleasure are repeated, and actions that induced punishment are avoided. To show how we determine evolutionarily stable strategies, we consider the simple case when $N = 10$, $p_A = .6$, and $c=0$ (bottom left panel of Figure 1). 
\par We first want to consider which regions of the graph correspond to positive or negative expected voting payoffs for both $A$ and $B$ supporters. Focusing on $A$ supporters (red curve), we know that their expected payoff for voting is exactly zero along this curve. If $q_A$ is unilaterally increased from a point on this curve, then candidate $A$ is more likely to win the election and so $A$ supporters are more likely to get the winning (positive when $c=0$) payoff for voting. Thus, to the right of the red curve, $A$ supporters have a positive expected payoff for voting (seen by the red plus in Figure 2). Analogously, when we decrease $q_A$ from a point on the red curve, we find that candidate $A$ is more likely to lose the election, and so $A$ supporters are more likely to get the negative payoff for voting when candidate $A$ loses. To the left of the red curve then, $A$ supporters have a negative expected payoff for voting (seen by the red minus in Figure 2). 
\par Similarly, we find that when $B$ supporters increase $q_B$ from a point on the blue curve, it will increase candidate $B$'s likelihood of winning, and $B$ supporters will have a positive expected payoff for voting. Below the blue curve, then, $B$ supporters will have a negative expected payoff for voting. These signs can also be seen in Figure 2.

\begin{figure}[htp]
\centering
\includegraphics[width=8cm]{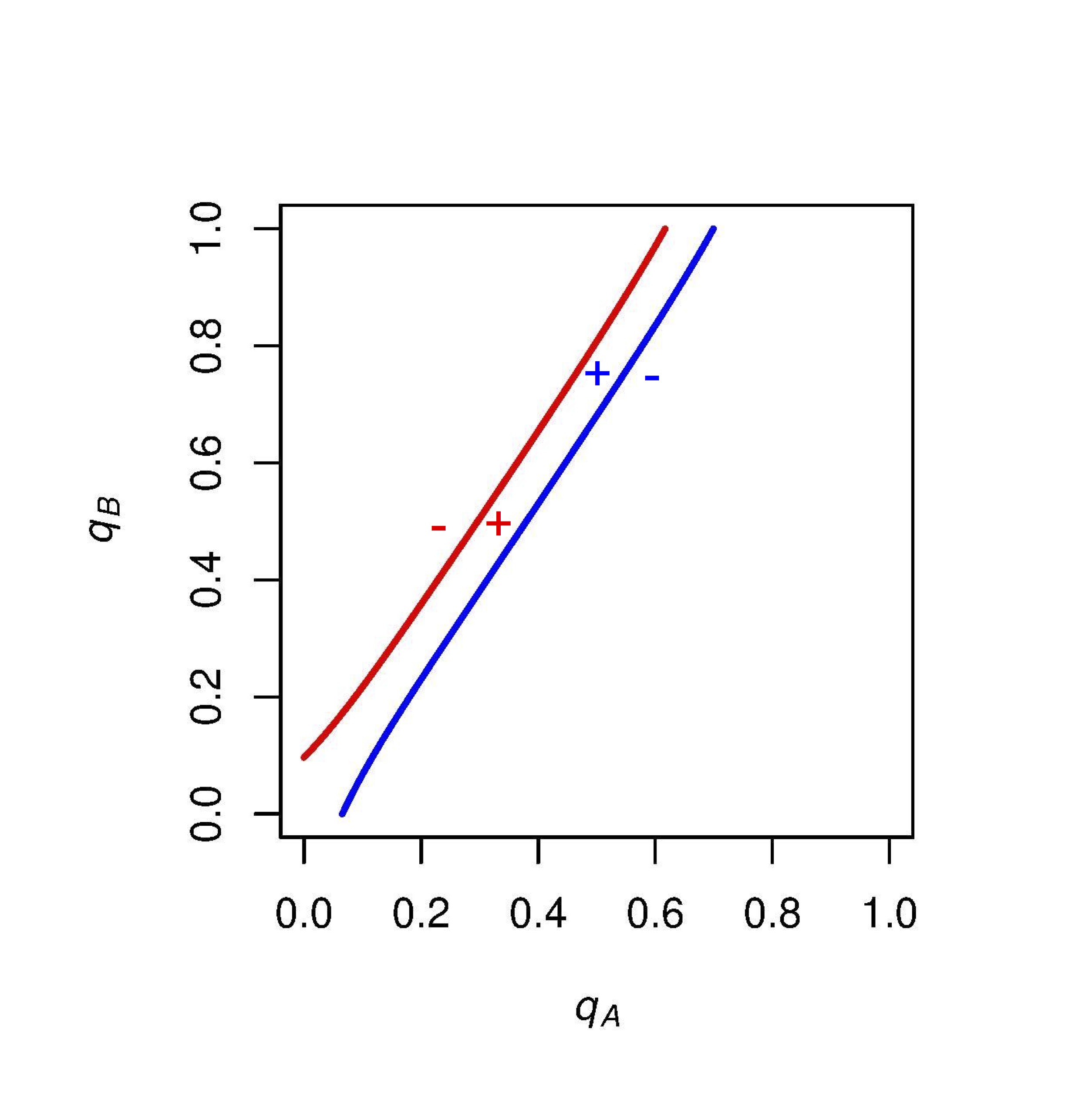}
\caption{Example graph with expected voting payoff signs. To the right of the red curve, the expected voting payoff for an $A$ supporter is positive, and to the left of this curve it is negative. Above the blue curve, the expected voting payoff for a $B$ supporter is positive, and below this curve it is negative. Parameters are set to $N = 10$, $p_A = .6$, and $c = 0$.}
\label{Fig2}
\end{figure}

\par In regions where the expected payoff for voting is positive, groups will want to increase the frequency with which they vote, and in negative regions, they will want to increase the frequency with which they abstain.
\par As noted earlier, voters are not rational utility maximizers; that is, they do not calculate which decisions yield the best outcomes. Rather, they start with a random choice, and then repeat actions that induced pleasure, and avoid actions that induced penalty. These random initial choices for $q_A$ and $q_B$ can lie in 3 different regions that have different implications for voter behavior.
\par If $A$ and $B$ supporters choose initial $q_A$ and $q_B$ values such that the initial state is above $A$ supporters' indifference curves, then in this region $A$ supporters get a negative expected payoff for voting, and $B$ supporters get a positive expected payoff for voting (Figure 3a). Consequently, $A$ supporters will decrease the likelihood with which they vote (leftward-pointing arrows in Figure 3a) and $B$ supporters will increase the likelihood with which they vote (upward-pointing arrows in Figure 3a). This process will iterate, with $q_A$ decreasing and $q_B$ increasing until turnout "runs away" to the stable state of no $A$ turnout and full $B$ turnout.
\par If, instead, the initial state is below $B$ supporters' indifference curve, then voting has a positive expected payoff for $A$ supporters and a negative expected payoff for $B$ supporters (Figure 3b). Thus, $q_A$ will increase and $q_B$ will decrease until we reach the other stable state of full $A$ turnout and no $B$ turnout. 
\par Finally, if the initial state is between these curves, then all individuals have a positive expected payoff for voting (Figure 3c). Both $A$ supporters and $B$ supporters will increase their voting probabilities until turnout breaks into one of the two original regions, and runs away to the respective stable state of either full $A$ turnout and no $B$ turnout or no $A$ turnout and full $B$ turnout. So whenever $c = 0$ and the indifference curves are configured in this way, the two stable states are full $A$ turnout and no $B$ turnout, or no $A$ turnout and full $B$ turnout. 

\begin{figure}[htp]
\centering
\includegraphics[width=\columnwidth]{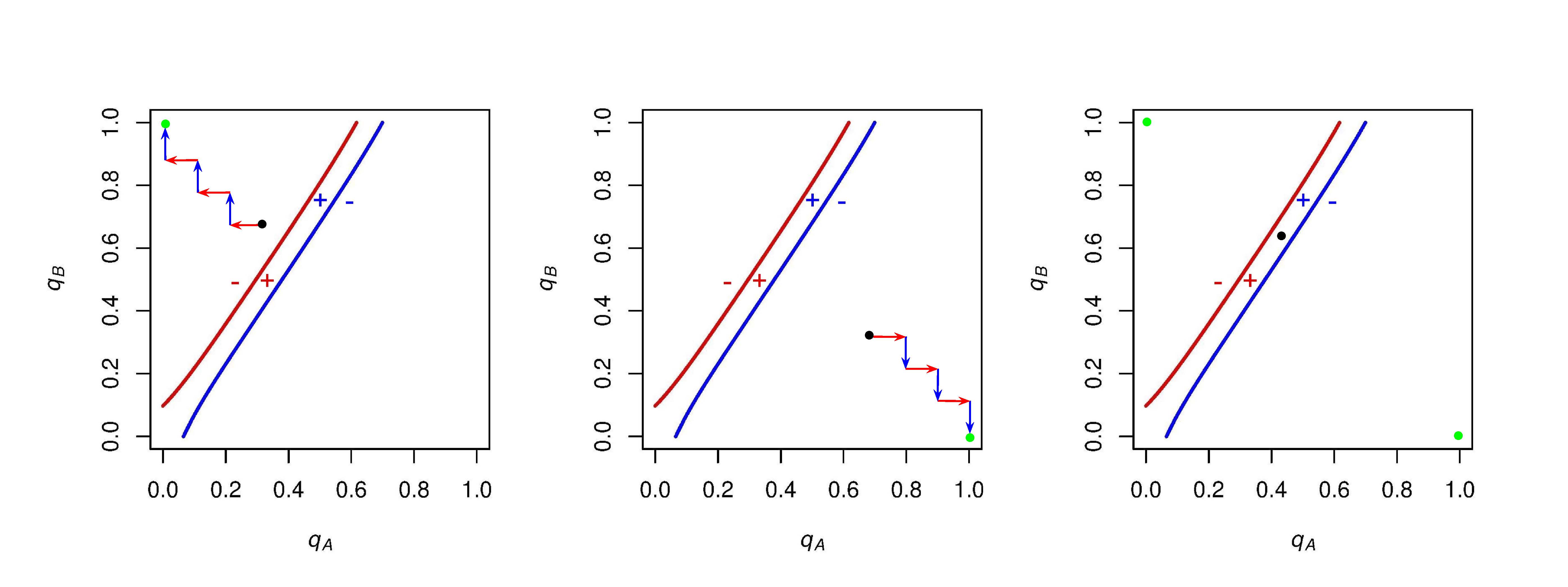}
\caption{Evolutionarily Stable Strategies when $N = 10, p_A = .6$, and $c = 0$. 3a (left panel): If $A$ supporters and $B$ supporters choose voting probabilities such that the initial state (represented by a black point) is in the top left region of the graph, then turnout will run away to the evolutionarily stable strategy (denoted by the green point) of $q_A=0, q_B=1$. 3b (middle panel): If the initial state is in the bottom right region of the graph, the evolutionarily stable strategy will be $q_A=1, q_B=0$. 3c (right panel): If the initial state is between these curves, both $q_A$ and $q_B$ will increase until the state breaks into one of the first two regions. This will lead to one of the two evolutionarily stable strategies from before.}
\label{Fig3}
\end{figure}

\par We now go on to analyze how these indifference curves and their associated stable states change as we alter $c$. Figure 4 shows the effect of increasing $c$ from 0 but below $1/(p_A - 0.5)$, the value for which the voting payoff for an $A$ supporter when candidate $A$ wins would become negative. We hold $N$ and $p_A$ constant at 10 and .6 respectively throughout this analysis in order to make our evolutionarily stable strategies more visually compelling, but the types of evolutionarily stable strategies are the same if we vary either $N$ or $p_A$ within the ranges of $c$ that we enumerate later. The effect of changing these parameters while holding $c$ constant is analyzed in Section 3.2. 

\begin{figure}[htp]
\centering
\includegraphics[width=\columnwidth]{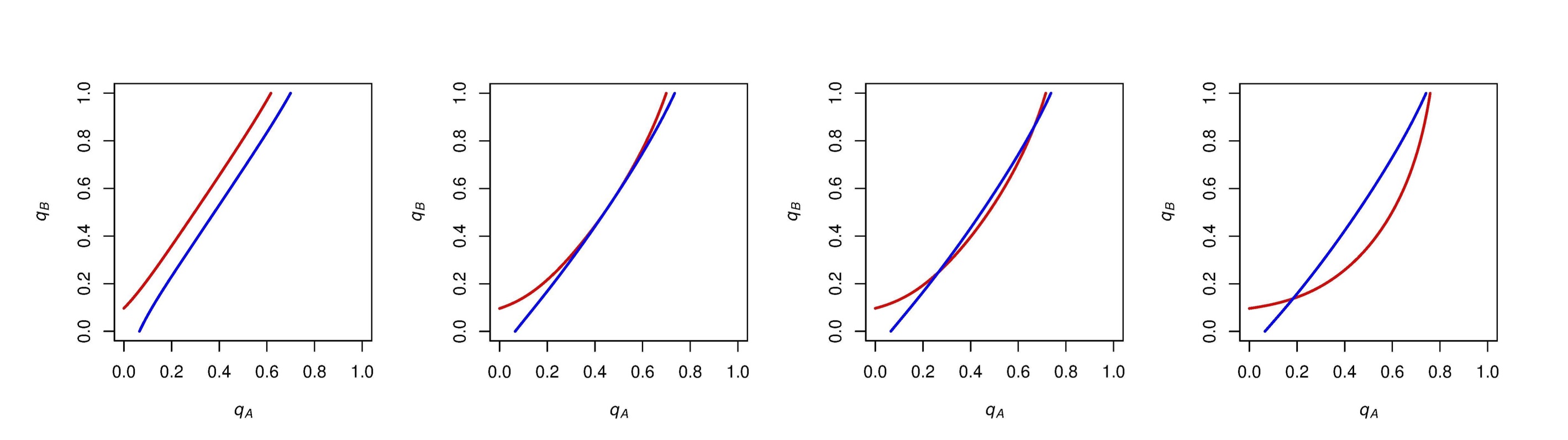}
\caption{The effect of increasing $c$ from zero (left-most panel) but below $1/(p_A - 0.5)$ (increasing from left to right). At first there is one mixed Nash when $A$'s indifference curve is tangent to $B$'s. As we increase $c$ further, the curves intersect twice, and as we increase $c$ beyond this point, there is just one low-turnout intersection.}
\label{Fig4}
\end{figure}

\par We now go on to analyze what evolutionarily stable strategies are rationalizable in the cases shown in Figure 4, when there exist intersections (and associated mixed-strategy Nash equilibria) between $A$ and $B$'s indifference curves. Figure 5 shows the results of this analysis.
\par We find that, although these curves have mixed-strategy equilibria, they are unstable: any infinitesimal deviation by either $A$ or $B$ supporters from these Nash probabilities would give rise to further deviation. In each of these cases, we find that turnout once again runs away to the same two stable states as before: either full $A$ turnout and no $B$ turnout, or no $A$ turnout and full $B$ turnout.

\begin{figure}[htp]
\centering
\includegraphics[width=\columnwidth]{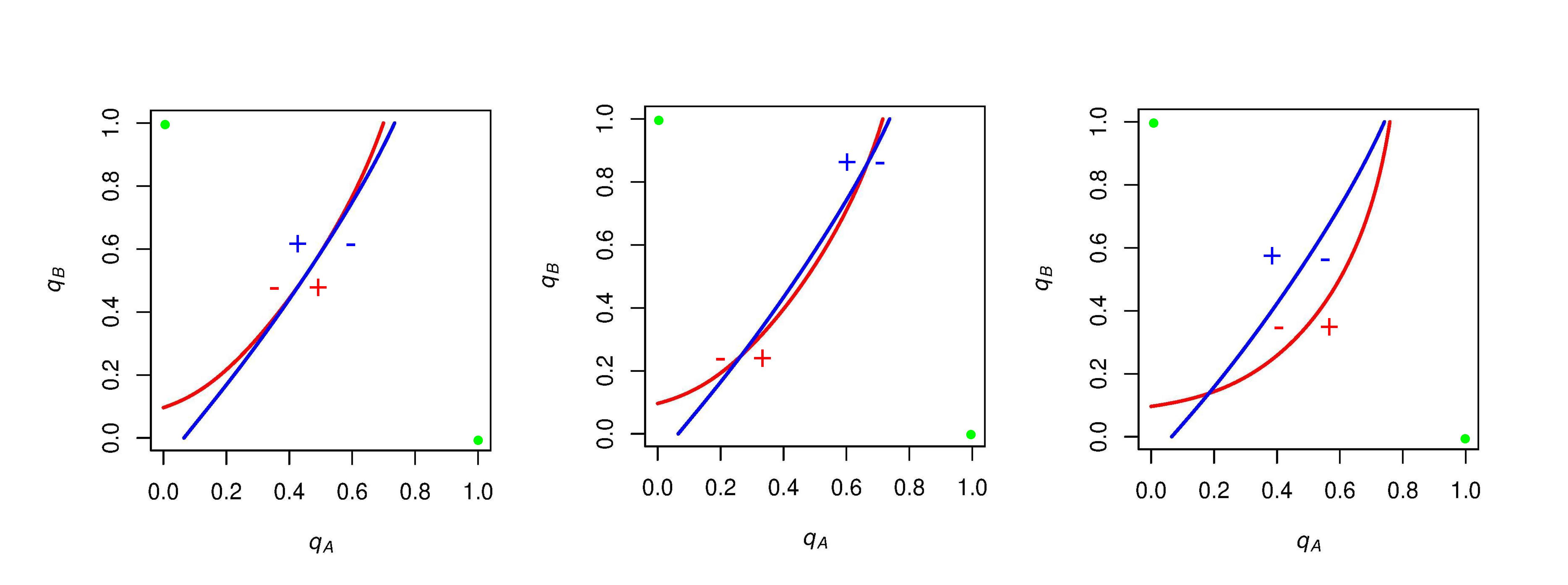}
\caption{Regardless of the number of intersections between $A$ and $B$'s indifference curves, when $0 \leq c \leq 1/(p_A - 0.5)$, the only evolutionarily stable strategies (represented by green points) that survive are $q_A = 1$ (full $A$ turnout) and $q_B = 0$ (no $B$ turnout) or $q_A = 0$ (no $A$ turnout) and $q_B = 1$ (full $B$ turnout).}
\label{Fig5}
\end{figure}

\par We now go on to analyze the effect that increasing $c$ beyond $1/(p_A - 0.5)$ will have on the configurations and respective turnouts of $A$ and $B$'s indifference curves. It is notable that once $c$ exceeds $1/(p_A - 0.5)$, the payoff for an $A$ supporter voting when candidate $A$ wins becomes negative. In this situation, the only way that an $A$ supporter can get a positive payoff for voting is if the candidates tie, and this individual's vote is pivotal in preventing a loss for candidate $A$. We first focus on cases when $1/(p_A - 0.5) < c \leq 2/(p_A - 0.5)$, as once $c$ exceeds $2/(p_A - 0.5)$, the payoff for an $A$ supporter voting when candidate $A$ wins becomes lower than -1, which is the payoff for an $A$ supporter voting when candidate $A$ loses. The effect of increasing $c$ between $1/(p_A - 0.5)$ and $2/(p_A - 0.5)$ is shown in Figure 6.

\begin{figure}[htp]
\centering
\includegraphics[width=\columnwidth]{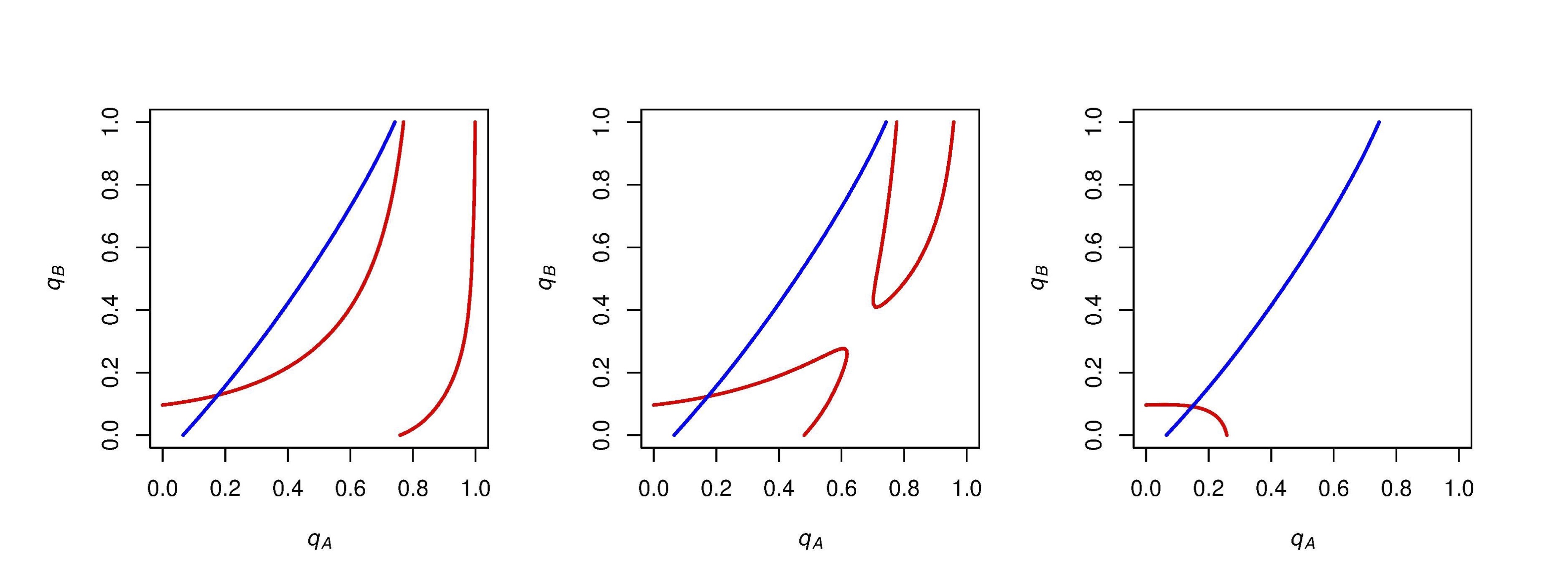}
\caption{The effect of increasing $c$ on the configuration $A$ and $B$'s indifference curves when $1/(p_A - 0.5) < c \leq 2/(p_A - 0.5)$. $c$ is increasing in panels from left to right.}
\label{Fig6}
\end{figure}

\par We go on to analyze evolutionarily stable strategies for these examples in Figure 7. It is noteworthy that $A$'s expected voting payoff is not always increasing in $q_A$ when $c$ is in this range. If $q_A$ is sufficiently high (and $q_B$ is sufficiently low), then $A$ supporters will be likely to win and accordingly get the negative payoff associated with voting and winning. This will compel $A$ supporters to abstain more and increase the likelihood of a tied election.
\par In each of these examples, it is clear that the stable state of no $A$ turnout and full $B$ turnout still exists, but now the stable state of full $A$ turnout and no $B$ turnout disappears. Instead, each of these cases has a stable state with $some$ $A$ turnout in expectation and no $B$ turnout. This expected $A$ turnout is decreasing in $c$ (left to right in Figure 7).
\par We find that $A$ supporters' ability to vote with these probabilities strictly between zero and one in equilibrium is supported by the fact that $A$ supporters always have a curve that is \emph{stable} when $c$ is in this range; that is, if $A$ supporters increase $q_A$ beyond (to the right of) this stable curve, they will get a negative expected payoff from voting and will abstain more by decreasing $q_A$, moving back towards the curve. If $q_A$ is decreased beyond (to the left of) this stable curve, $A$ supporters will get a positive expected payoff from voting, and will vote more by increasing $q_A$ back towards this curve.

\begin{figure}[htp]
\centering
\includegraphics[width=\columnwidth]{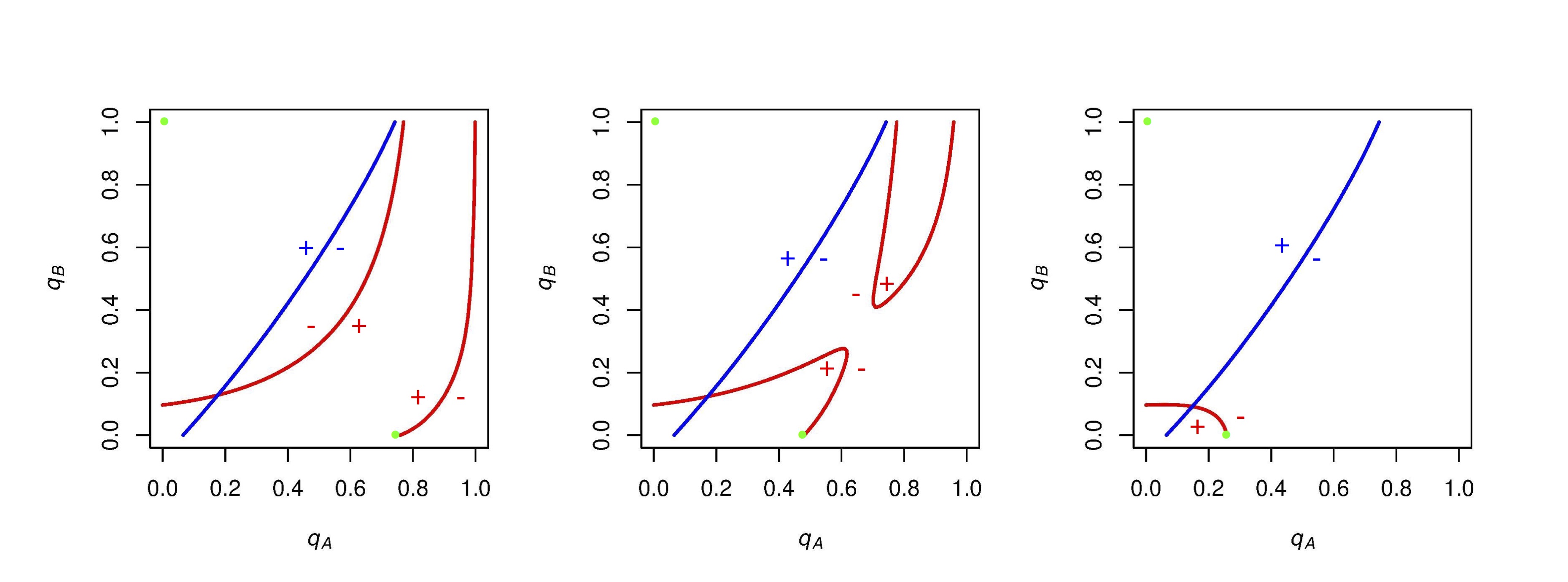}
\caption{Evolutionarily stable strategies as we increase $c$ when $1/(p_A - 0.5) < c \leq 2/(p_A - 0.5)$. Stable states either have no $A$ turnout and full $B$ turnout or $some$ $A$ turnout in expectation and no $B$ turnout. Expected $A$ turnout is decreasing in $c$ for the latter stable state.}
\label{Fig7}
\end{figure}

\par In Figure 8 we analyze what happens when $c$ exceeds $2/(p_A - 0.5)$. Remember that when $c$ takes on such values, $A$ supporters' voting payoff when candidate $A$ wins is now $less$ than $A$ supporters' voting payoff when candidate $A$ loses (-1). We find that whenever $c$ exceeds this value, $A$ and $B$'s indifference curves no longer intersect, and the only stable state that remains has no $A$ turnout and full $B$ turnout.

\begin{figure}[htp]
\centering
\includegraphics[width=8cm]{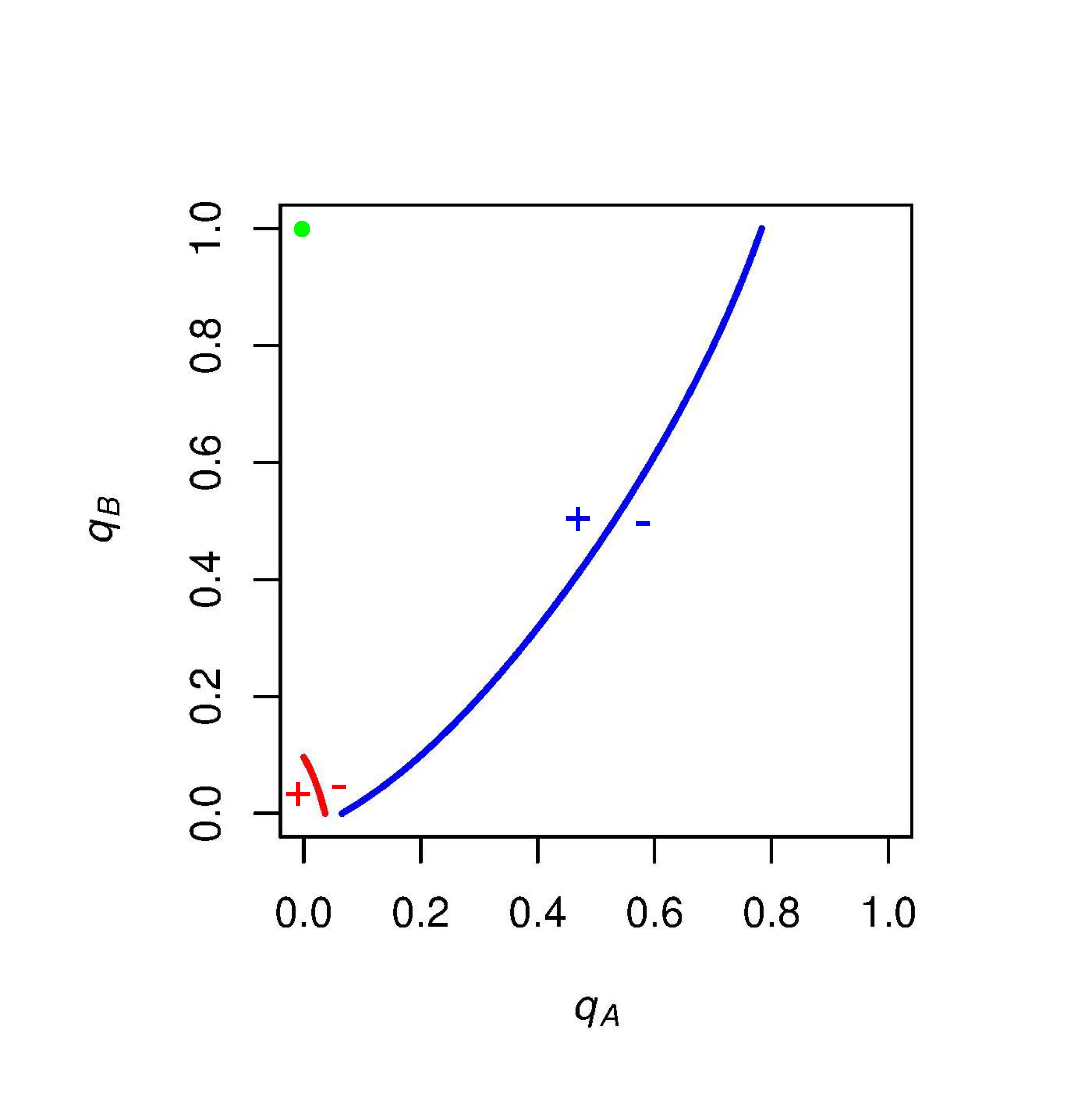}
\caption{Evolutionarily stable strategy when $c > 2/(p_A - 0.5)$. The only stable state has no $A$ turnout and full $B$ turnout.}
\label{Fig8}
\end{figure}

\par To summarize, when $0 \leq c \leq 1/(p_A - 0.5)$, there are two stable states: no $A$ turnout and full $B$ turnout, or full $A$ turnout and no $B$ turnout. When $1/(p_A - 0.5) < c \leq 2/(p_A - 0.5)$, there are also two stable states: no $A$ turnout and full $B$ turnout, or $some$ $A$ turnout (in expectation) and no $B$ turnout. Lastly, when $c > 2/(p_A - 0.5)$, the only stable state that remains has no $A$ turnout and full $B$ turnout. We now go on to consider what negative levels of $c$ might mean qualitatively, and what implications relaxing this assumption might have for voter behavior.

\subsection{Turnout when $c \leq 0$}
\par It is interesting to consider whether having a negative value for $c$ makes sense in respect to human behavior. This would mean that those who perceive themselves likely to win (in our case, those in the majority) get an additional benefit for voting when their candidate wins, and those who perceive themselves unlikely to win (in our case, those in the minority) get a penalty on their voting payoff when their candidate wins. 
\par Rather than having an \emph{underdog effect,} wherein people in the minority enjoy their victory more than those in the majority, we now have a \emph{will-of-the-people effect.} In this type of world, voters in the majority get a benefit when their candidate wins for supporting the election of a candidate that more people prefer. Voters in the minority get a penalty when their candidate wins as they may feel some guilt for electing a candidate that is not as well-supported (leaving this voter feeling somewhat selfish).
\par With this framework in mind, we go on to analyze what stable states arise from different negative values of $c$, finding relatively symmetric results to the positive values of $c$. Figure 9 shows the results of decreasing $c$ from zero but not beyond $1/(0.5-p_A)$, the point at which $B$ supporters' voting payoff when candidate $B$ wins becomes negative. As we decrease $c$ (left to right in Figure 9), we find that, although we get a varying number of Nash equilibria, we end up with the two stable states of full $A$ turnout and no $B$ turnout or no $A$ turnout and full $B$ turnout that we found when $0 \leq c < 1/(p_A-0.5)$.

\begin{figure}[htp]
\centering
\includegraphics[width=\columnwidth]{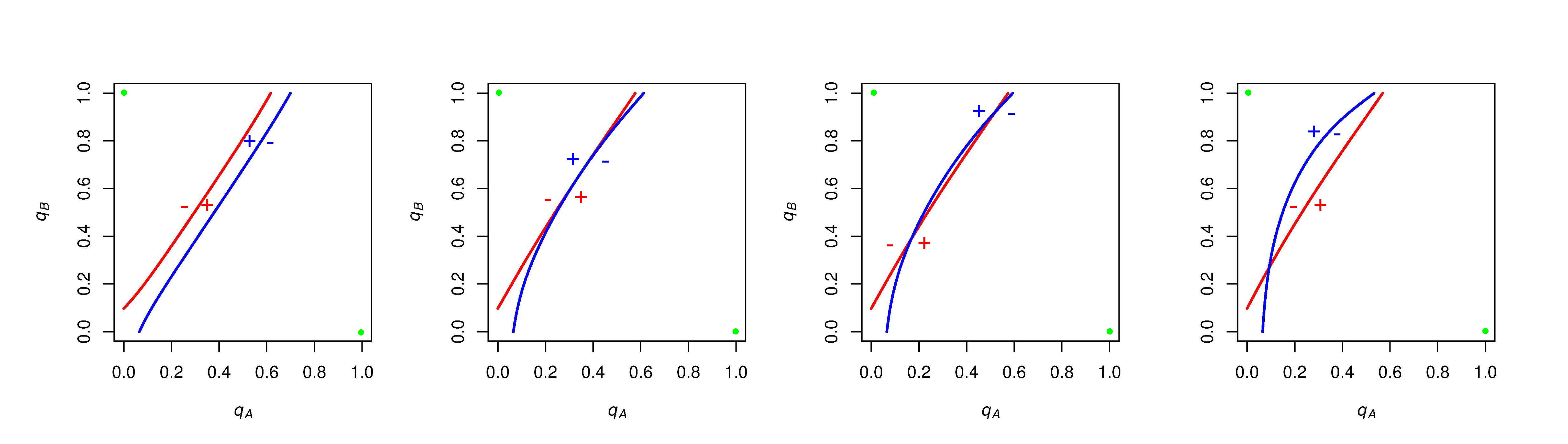}
\caption{The effect of decreasing $c$ (left to right) from zero but not beyond $1/(0.5-p_A)$. Regardless of the number of Nash equilibria, there are only two stable states: full $A$ turnout and no $B$ turnout or no $A$ turnout and full $B$ turnout.}
\label{Fig9}
\end{figure}

\par Similar to the positive $c$ case, we find that our evolutionarily stable strategies change once $c$ exceeds $1/(0.5-p_A)$. Beyond this point, the voting payoff for a $B$ supporter when candidate $B$ wins becomes negative, meaning that $B$ supporters can only get a positive payoff from voting if the result of the election is a tie. 
\par This leads to a symmetric finding to the positive $c$ cases: when $2/(0.5 - p_A) \leq c < 1/(0.5 - p_A)$, $B$ supporters have an indifference curve that is $stable$. If $B$ supporters deviate from this curve by increasing $q_B$, they will receive a negative payoff for voting, and will abstain more, decreasing $q_B$ towards this curve. If they deviate from this curve by decreasing $q_B$, they will receive a positive payoff from voting, will vote more, and increase $q_B$ back towards this curve.
\par This allows for equilibria in which supporters from one bloc vote with a probability strictly between zero and one. Once again, we find that this voting probability is decreasing in the magnitude of $c$ (in the negative direction, in this case). The results of this analysis can be seen in Figure 10.

\begin{figure}[htp]
\centering
\includegraphics[width=\columnwidth]{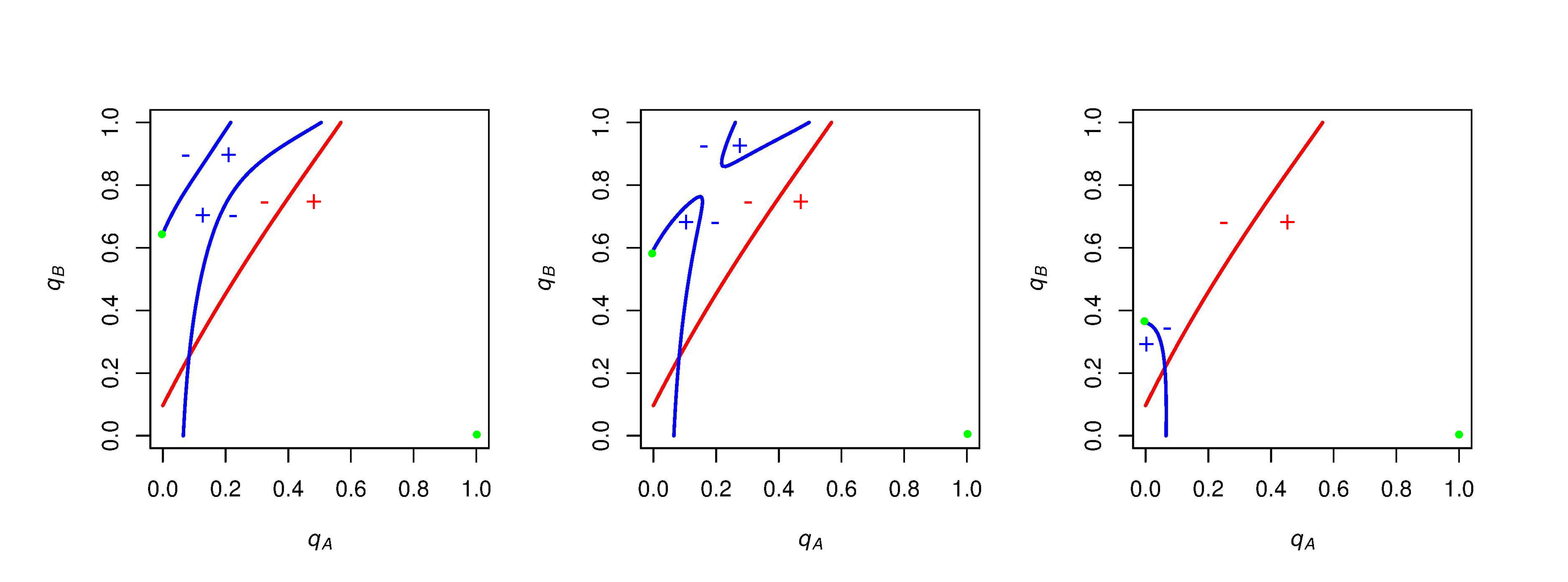}
\caption{The effect of decreasing $c$ (left to right) when $2/(0.5 - p_A) \leq c < 1/(0.5 - p_A)$. In each of these cases, there are two stable states: full $A$ turnout and no $B$ turnout, or no $A$ turnout and $some$ $B$ turnout in expectation. Expected $B$ turnout for the latter case is decreasing in the magnitude of $c$.}
\label{Fig10}
\end{figure}

\par When we decrease $c$ beyond $2/(0.5 - p_A)$, $B$ supporters' payoff for voting when candidate $B$ wins becomes less than the payoff for voting when candidate $B$ loses (-1). In this case, as with its positive analog, $A$ and $B$'s indifference curves no longer intersect, and the stable state with partial turnout in expectation disappears (Figure 11).

\begin{figure}[htp]
\centering
\includegraphics[width=8cm]{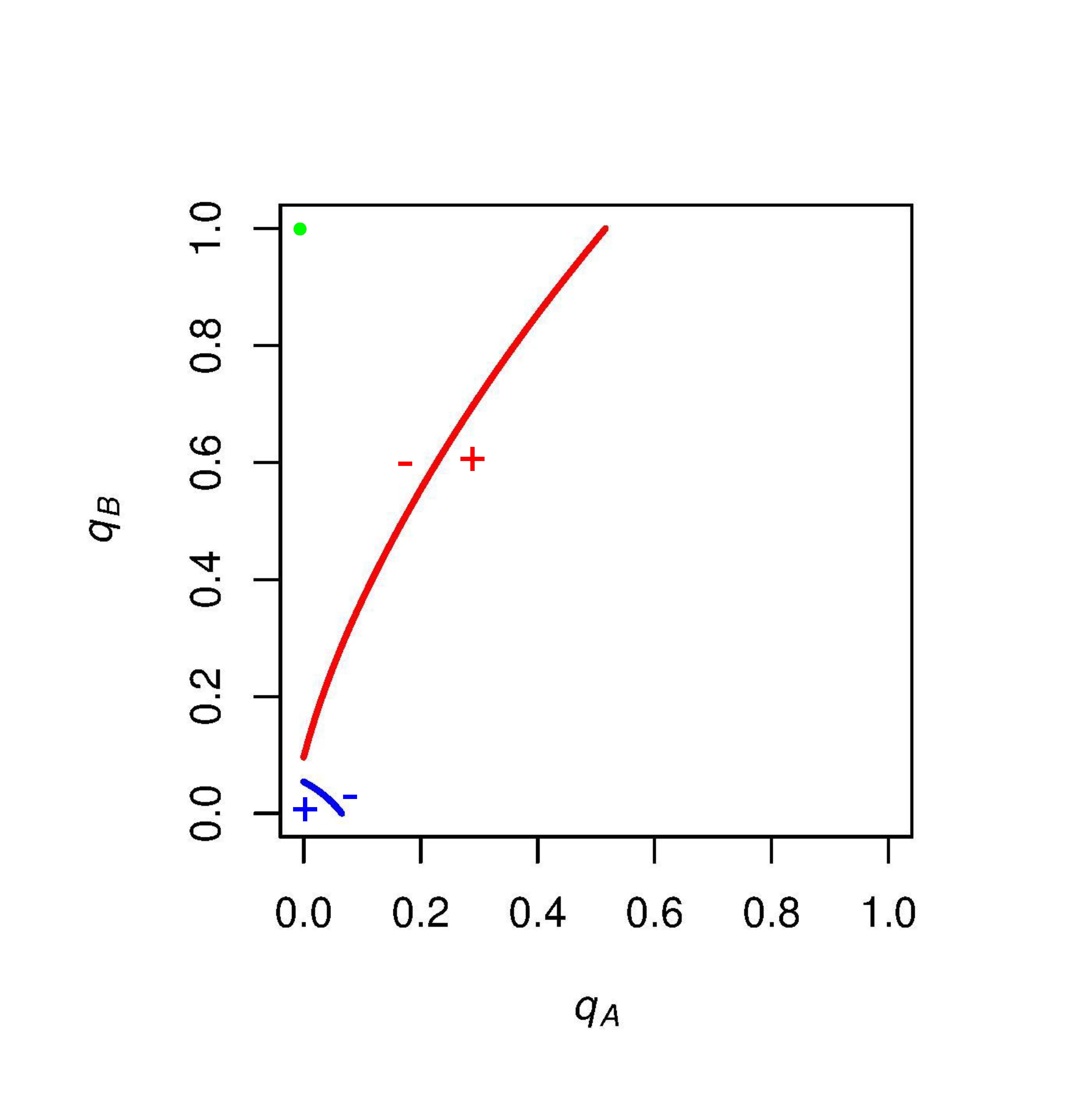}
\caption{An example graph for when $c < 2/(0.5 - p_A)$. When $c$ is less than this value, the indifference curves no longer intersect, and the only stable state has full $A$ turnout and no $B$ turnout.}
\label{Fig11}
\end{figure}

\par At this point we have covered all of the different ranges for $c$ and analyzed what evolutionarily stable stategies exist in each case. A summary of this analysis can be found in Table 1.

\begin{table}[htp]
\caption{Evolutionarily Stable Strategies Summary}
\centering
\begin{tabular}{|c|c|c|}
\hline
$c$ Range & $ESS$ 1 & $ESS$ 2 \\\hline
$c > 2/(p_A-0.5)$ & $q_A = 0, q_B = 1$ & \\
\hline
$1/(p_A-0.5) < c \leq 2/(p_A-0.5)$ & $q_A = 0, q_B = 1$ & $0<q_A<1, q_B = 0$\\
\hline
$1/(0.5-p_A) \leq c \leq 1/(p_A-0.5)$ & $q_A = 0, q_B = 1$ & $q_A = 1, q_B = 0$ \\
\hline
$2/(0.5-p_A) \leq c < 1/(0.5-p_A)$ & $q_A = 1, q_B = 0$ & $q_A=0, 0<q_B<1$ \\
\hline
$c < 2/(0.5-p_A)$ & $q_A = 1, q_B = 0$ & \\ [1ex]
\hline
\end{tabular}
\end{table}

\subsection{Comparative Statics}
\par Another thing worth considering is the effect that changing either the size of the population ($N$) or the partisan spread ($p_A$ vs.\ $p_B$) would have on voter behavior. Remember that we have thus far focused on the simplistic case of a 6 vs.\ 4 person election for ease of analysis. While we have analyzed the effect of varying $c$ on turnout for this particular case, we want to see how varying these other parameters might affect voter behavior.
\par Table 1 shows that many of our stable states are of the \emph{all or nothing} variety, with either full turnout or no turnout from the supporters of each party. We find that these stable states remain regardless of changes to $N$ and $p_A$; whenever $c$ is within the intervals outlined in Table 1, the \emph{all or nothing} stable states are not altered by changes in either the number of individuals in the electorate or the partisan spread of these individuals.
\par With this in mind, we turn to the cases when one of the parties votes with a probability strictly between zero and one. We focus on the case when $1/(p_A-0.5) < c \leq 2/(p_A-0.5)$ and there exists an evolutionarily stable strategy in which $0<q_A<1$, but we will discuss the comparative statics for the negative analog of this case as well. Figure 12 provides a visual representation of how this $ESS$ changes as we vary $N$ and $p_A$.

\begin{figure}[htp]
\centering
\includegraphics[width=12cm]{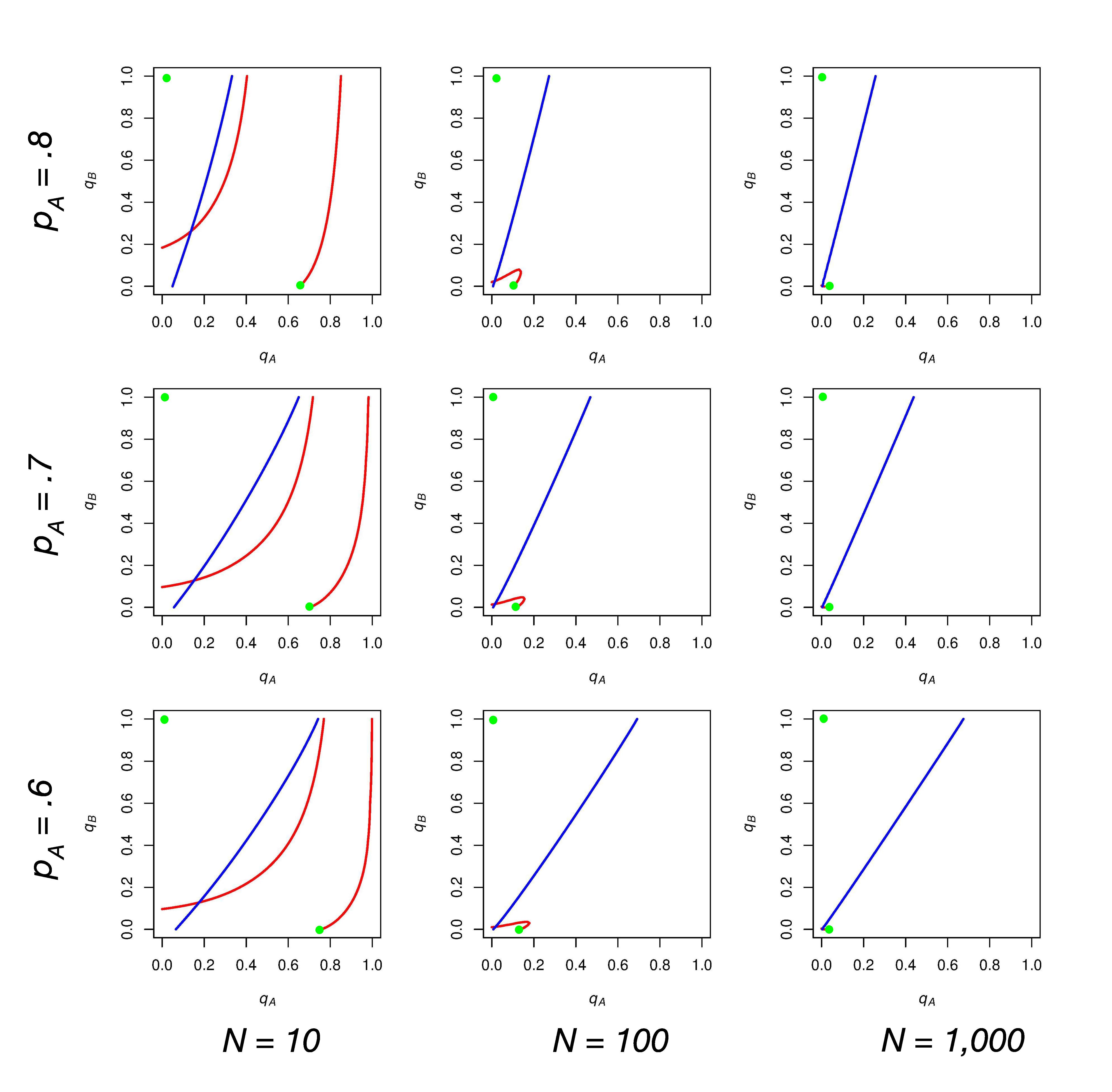}
\caption{The effect of varying $N$ and $p_A$ while holding $c$ constant at $10,001/(10,000(p_A-0.5))$. The evolutionarily stable strategy when $A$ supporters vote with a probability strictly between zero and one has expected turnout that is decreasing in $N$ as well as $p_A$, $ceteris$ $paribus$.}
\label{Fig12}
\end{figure}

\par We find that both $N$ and $p_A$ effect expected turnout for these equilibria. The more striking influence is this \emph{size effect} predicted by our model, wherein expected turnout is decreasing (quite rapidly) in $N$. 
\par This makes sense when we consider the range of $c$ that we are considering. When $1/(p_A-0.5) < c \leq 2/(p_A-0.5)$, the only way $A$ supporters can get a positive payoff for voting is if the result of the election is a tie. In order to drive turnout away from zero in expectation for $A$ supporters, then, a tie must be sufficiently likely to occur; as we increase $N$ while holding $c$ and $p_A$ constant, the probability of the election resulting in a tie approaches zero, and thus $A$ supporters become unable to vote at high rates. Focusing on the bottom row of Figure 12, one can imagine how the likelihood of a tie may change from a 6 vs.\ 4 person election to a 60 vs.\ 40 person election to a 600 vs.\ 400 person election.
\par Furthermore, we find this size effect to hold true when looking at the negative analog of this case. When $2/(0.5-p_A) \leq c < 1/(0.5-p_A)$ and we focus on the evolutionarily stable strategy in which $0<q_B<1$, expected turnout is also decreasing in $N$.
\par Back to the positive case, when viewing the impact that varying $p_A$ has on expected turnout, it is less dramatic than the impact of varying $N$. It is nonetheless the case that expected turnout is decreasing in $p_A$. Let us consider why this is so.
\par If we hold $N$ and $c$ constant while increasing $p_A$, we are increasing the partisan spread, and making candidate $A$ more likely to win, all else equal. As $A$ supporters prefer a tie to candidate $A$ winning (as their vote is pivotal), at high $p_A$'s, $A$ supporters will have a decrease in the $q_A$-cutpoint at which their expected payoff for voting switches from positive to negative. This cutpoint (when $q_B$ is equal to zero) is exactly the evolutionarily stable strategy we are discussing. Therefore, we find expected turnout to be decreasing in $p_A$ for this type of equilibrium.
\par Once again, we consider whether this comparative static holds for the negative analog of this case. In the negative $c$ scenario, it is $B$ supporters who will only get a positive payoff from voting if there is a tie. If we consider the $q_B$-cutpoint at which $B$ supporters' expected payoff switches from positive (tie is sufficiently likely) to negative (candidate $B$ winning is sufficiently likely), we should find this point increasing in $p_A$. When $B$ supporters comprise only a very small proportion of the electorate (large $p_A$), then even at high $q_B$ levels, it is less likely that candidate $B$ will win in comparison to when $p_A$ is smaller. Thus, we find the opposite $p_A$ effect in the negative case: expected turnout is increasing in $p_A$.
\par While we do not find a consistent effect of partisan split on expected voter turnout, we do find that, whenever $N$ affects our equilibrium state, expected voter turnout monotonically decreases in $N$. We go on to see if this is consistent with real voter behavior.
\section{Data}
\par In general, this size effect, in which turnout and electorate size are negatively correlated, has been found to be consistent with voter behavior in a myriad of contexts. While many cite this as evidence that Downs' formulation of the instrumental voter is correct, others have interpreted the somewhat weak but consistent correlation between electorate size and turnout as evidence that the instrumentality of one's vote is only $part$ of the motivation for voting, an interpretation in agreement with out model.
\par For example,~\cite{levine2007paradox} summarizes the results of an experiment conducted in order to examine how voter turnout is affected by different variables, electorate size among them. Using \emph*{electorate sizes} no larger than 51, they find strong evidence that size and turnout are negatively correlated. Levine and Palfrey still find, though, that the size effect cannot entirely explain voting if we consider the solely instrumental voter; rather, they propose that a blend of rationality models may be necessary to fully explain voter behavior.

\begin{figure}[htp]
\centering
\includegraphics[width=8cm]{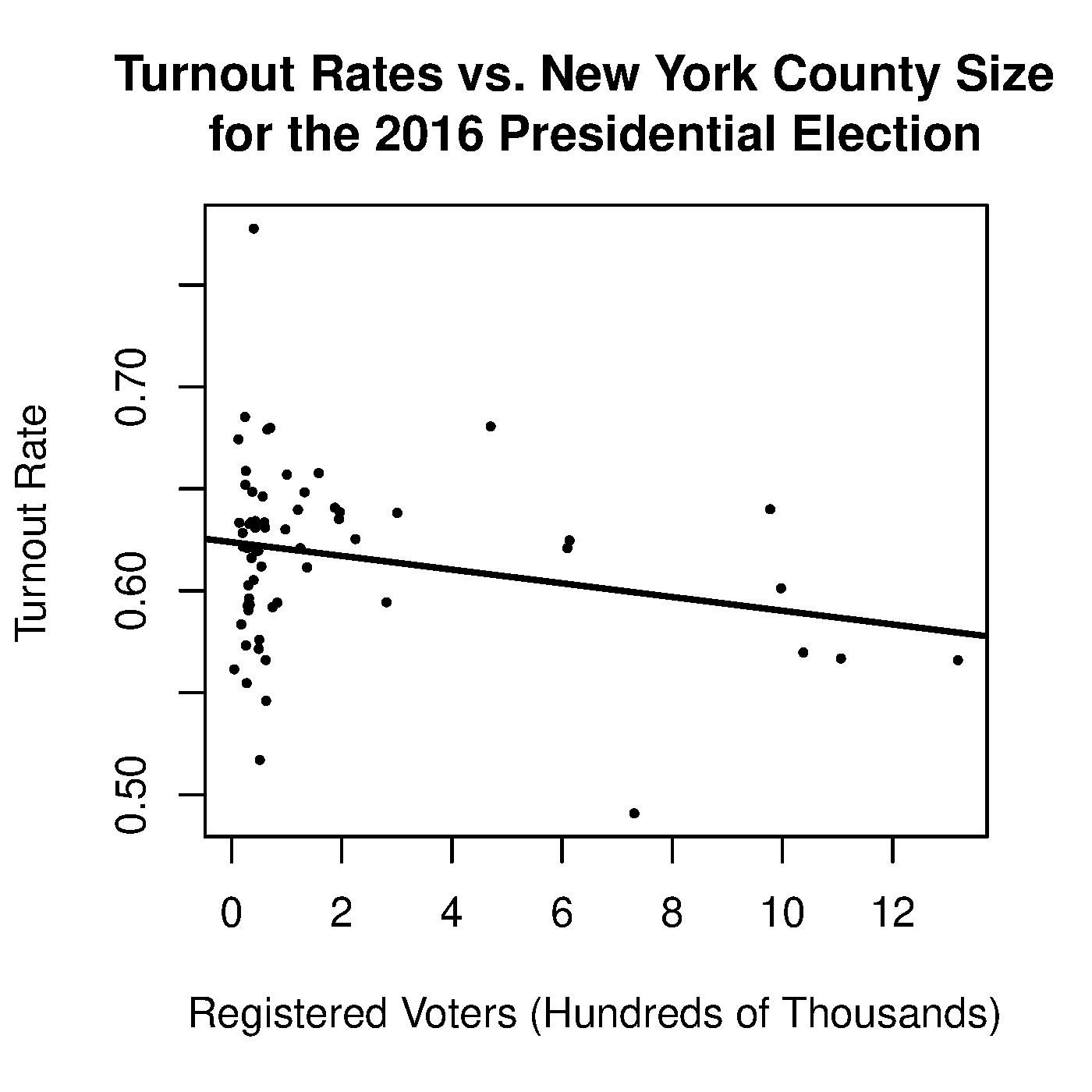}
\caption{We find that turnout is decreasing in $N$, similar to the comparative statics predicted by our model. We use the number of registered voters as reported by the NYS Board of Elections and the vote count as reported by OpenDataSoft. Turnout is calculated as the number of votes in the county divided by the number of registered voters in the county. Even when the probability of a tie is quite small, this size effect seems to persist in voter behavior.}
\label{Fig13}
\end{figure}

\par Observational studies support this size effect as well.~\cite{breux2017fewer} finds that electorate size is the most important factor in motivating turnout in the municipal elections they study, but that this impact is more prevalent in relatively small electorates. 
\par Still, even in elections with large electorates (gubernatorial elections in this case), the negative correlation between electorate size and turnout persists, although it is relatively weak~\cite{barzel1973act}. This is an interesting finding, as our model predicts that this size effect becomes relatively negligible (as expected turnout approaches zero quickly) as $N$ gets large. As the probability of a tie becomes staggeringly small in large electorates, we would expect turnout to decrease almost imperceptibly in $N$ for large electorates.
\par We analyze a dataset with smaller $N$'s than gubernatorial races- the New York State turnout for the 2016 presidential election by county (Figure 13). Plotting a least squares regression line over these points, we find that the relationship between voter turnout and electorate size has the same sign that our model predicts. This relationship is somewhat weak, again suggesting that individuals do not only consider whether or not their vote will be pivotal when they make the decision of whether or not to vote.
\par Still, it is notable that this relationship between electorate size and turnout persists even when we are dealing with electorates that our model predicts would have expected turnout rates extremely close to zero. In an attempt to reconcile this as well as another common turnout trend with our model, we go on to propose some possible extensions to our base model.

\section{Extensions}
\par Although our model was successful in the direction of its comparative static predictions, it is certainly not without its shortcomings. Our model still underestimates turnout for large electorates, and does not predict any stable states with expected turnout from both parties. While we believe that the formulation of voter motives posited by our base model is powerful, allowing for more variation in our parameters might lead to turnout predictions that are more descriptively accurate.
\par First, we observe the effect of relaxing the assumption that the partisan split of the electorate is common knowledge, as this is more often than not the case in the real world. A variety of factors can lead to misperceptions regarding the partisan split of an electorate. Some people are misguided about this from a lack of information, some from conflicting information, and some from the spread of false information, an issue ever so present in the most recent election cycles.
\par If we let $\tilde{p_A}$ equal the $perceived$ proportion of the population that supports candidate $A$, with $\tilde{p_B}$ defined analogously, we can analyze how these perceptions can impact voter turnout. Consistent with our finding from earlier, in the $ESS$ in which $0<q_A<1$, we find that turnout is decreasing in $p_A$. Thus, allowing $c$ to be a consistent function of $\tilde{p_A}$, we find that turnout is higher in expectation whenever $\tilde{p_A} < p_A$. We also found that, for the $ESS$ in which $0<q_B<1$, turnout is increasing in $p_A$, and so is decreasing in $p_B$. In this case, expected turnout  is higher whenever $\tilde{p_B} < p_B$.
\par This finding is both descriptive and prescriptive. As we've noted, people often vote at higher rates than our model predicts; however, we find that allowing misperceptions about the partisan split of a population might provide an explanation for behavior that is seemingly irrational. If individuals in the majority feel that the partisan split is closer than in reality it is, this will compel them to turn out at higher rates than they would if they knew the true distribution of preferences in the electorate. If individuals in the $minority$ think that a win is not likely, this will motivate more of these individuals to vote as they know every vote for their candidate is more vital to preventing a loss.
\par As a disclaimer, we do not advocate the spread of false information by politicians or anyone for that matter. However, we do find that bending the perceptions of a support base may have tangible impacts on voter turnout. If a candidate has a majority of the support in an election (particularly a large majority), then she runs the risk of her supporters free-riding and abstaining while a fraction of supporters vote and prop up the possibility of a tied election. Thus, we find that it may be beneficial for this candidate to downplay the majority support she holds when talking to her electorate. By convincing supporters that the race is closer than it actually is, she can avoid this free-rider problem and be more likely to secure her (arguably well-deserved) victory.
\par On the other side of this, there lies a strategy that a candidate with minority support might find beneficial for spurring turnout from his support base. When a candidate has minority support, it will behoove him to downplay the size of this support. This may seem counterintuitive, as one would think that emphasizing how close the race is would motivate turnout. However, our model finds that whenever a candidate overestimates the magnitude of one's support to his base (implementing a $\tilde{p_B}$ that is larger than $p_B$), he runs the risk of his supporters abstaining more. A candidate in this scenario would find it in his best interest to downplay the size of his support, compelling supporters to view their vote as more integral to preventing the loss of their candidate of choice.
\par While we offer the misperception of $p_A$ and $p_B$ as a possible solution to the over-voting we see in electoral behavior, we have yet to account for the fact that, in electoral politics, turnout rates are almost always strictly between zero and one for supporters of $both$ candidates. To account for this, we relax the assumption that the electorate for a given election has a universal $c$. If we allow for a (potentially) different $c_j$ for supporters of the different candidates, it is possible to obtain types of evolutionarily stable strategies that were not possible in the base model.  
\par For example, consider a case when $c_A$ is positive and $c_B$ is negative. Qualitatively, this means that $A$ voters receive a penalty for being in the majority when candidate $A$ wins (as this outcome seems inevitable anyways), and $B$ voters receive a penalty for being in the minority when candidate $B$ wins (as they help elect a candidate that is not well-supported by the electorate). Both blocs have voting payoffs when their candidate of choice wins that are decreasing in $p_A$.
\par When we set $c_A$ and $c_B$ to levels such that the payoff for voting when one's candidate of choice wins is less than the payoff for abstaining but greater than the payoff for voting when one's candidate of choice loses (between 0 and -1), we can find evolutionarily stable strategies that could not be obtained using our base model alone. Figure 14 shows one such case.

\begin{figure}[htp]
\centering
\includegraphics[width=8cm]{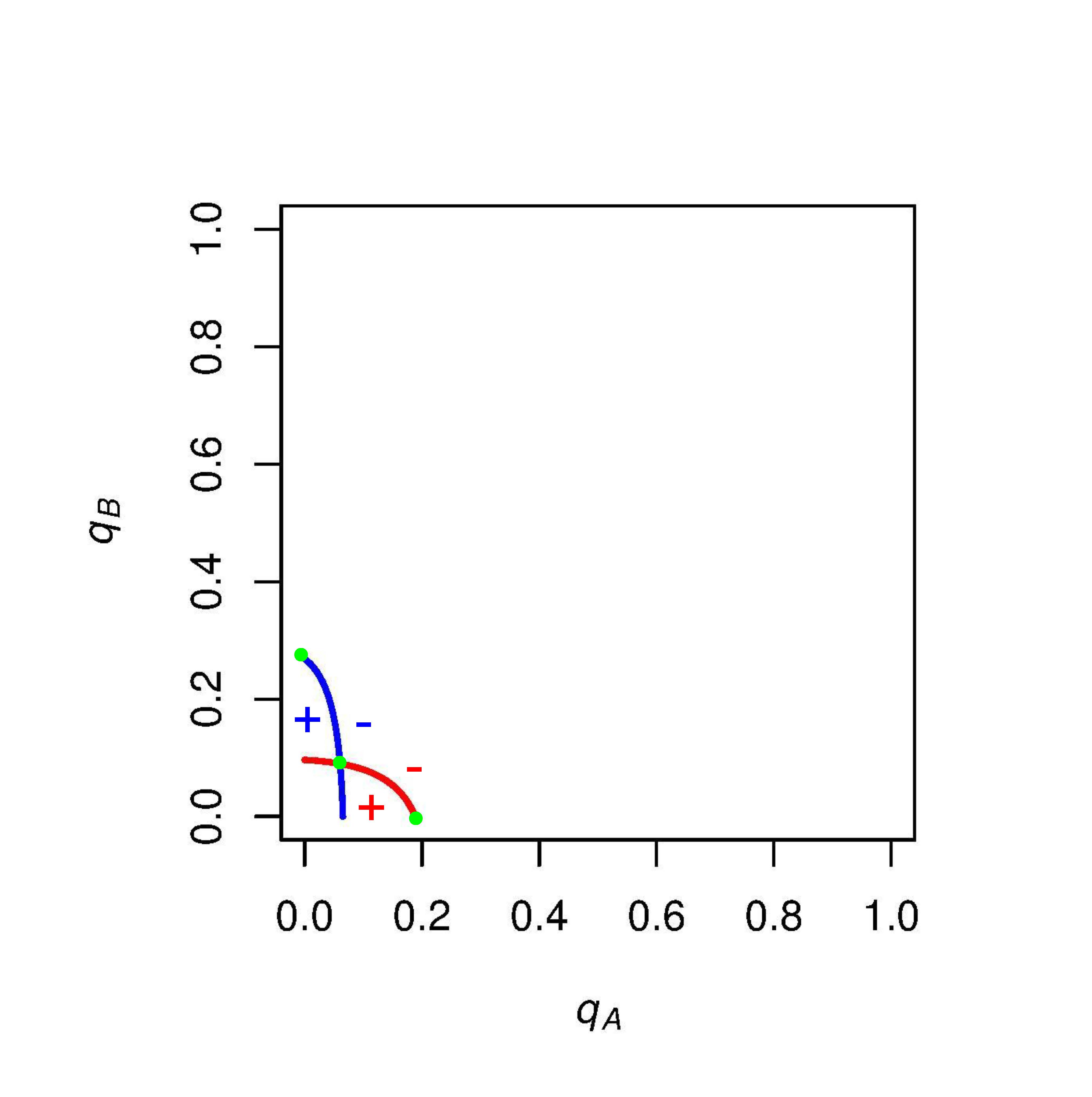}
\caption{Evolutionarily stable strategies when we allow $c$ diversity in our model. In this simple case ($N = 10$, $p_A = .6$, $c_A = 12$, $c_B = -12$), we find three $ESS$'s. Most notably there exists an $ESS$ with positive expected turnout from both parties.}
\label{Fig14}
\end{figure}

\par Based on these new parameters, our model predicts three distinct evolutionarily stable strategies: one in which $0<q_A<1$ and $q_B=0$, one in which $q_A=0$ and $0<q_B<1$, and one in which $0<q_A<1$ and  $0<q_B<1$. This third $ESS$ is notable because it reconciles our model with real electoral behavior. In nearly all elections, supporters of each party vote with turnout rates strictly between zero and one; while our base model is inconsistent with this finding, introducing $c$ diversity between supporters of different candidates into our payoff structure brings our model in accordance with such behavior.

\section{Conclusion}

 \par While many models have attempted to overcome Downs' original formulation of the voter problem, our model blends instrumental and expressive voting theories, allows for learning in the electorate, and introduces partisan asymmetry in a way that other models do not. We introduce a new variable, $c$, to account for the extent to which a given population does or does not consider the \emph{underdog effect} when making the decision to vote for one's candidate of choice or abstain.
 
 \par We find that changes in this variable compel different types of evolutionarily stable strategies. Depending on the range of $c$, our model predicts either one party turning out to vote in full and the other party abstaining in full, or one party partially turning out to vote in expectation and the other party abstaining in full. In the latter case, expected turnout is decreasing in electorate size as well as the magnitude of the party that is expected to vote. 
 
 \par This size effect is well-documented in turnout literature, with turnout decreasing in electorate size even when the probability that one's vote is pivotal is extremely small. An analysis of turnout in New York State counties from the 2016 presidential election agrees with this finding.
 
 \par While our model predicts turnout that is smaller than we often see in large electorates, relaxing some of our preliminary assumptions may bring our model more in accordance with true voter behavior. Specifically, we find that when individuals have misperceptions about the partisan split of an electorate (and, by proxy, the likelihood of affecting the outcome), turnout can be higher than our base model predicts. Furthermore, we find that, while our base model does not predict positive turnout from both support blocs in equilibrium (a phenomenon common in electoral politics), introducing $c$ diversity into our model allows this to exist. We find the incorporation of both partisan misperception and $c$ diversity into our model to be especially promising in explaining and predicting voter turnout moving forward.
 
\section*{Author contributions}
C.G. \& F.F. conceived the model and performed theoretical analysis, C.G. made figures and did real voter turnout data analysis and wrote the first draft, and C.G. \& F.F. contributed to the revision of the manuscript.  

\section*{Acknowledgements}
C.G. is grateful for generous financial support by the Byrne Scholars Program, Sophomore Research Scholarship, and the Mellam Family Foundation Research Award at Dartmouth. F.F. is supported by a Junior Faculty Fellowship awarded by the Dean of the Faculty at Dartmouth and also by the Bill \& Melinda Gates Foundation (award no. OPP1217336), the NIH COBRE Program (grant no. 1P20GM130454), the Neukom CompX Faculty Grant, the Dartmouth Faculty Startup Fund, and the Walter \& Constance Burke Research Initiation Award.

\end{document}